\documentclass{article}

\usepackage[english]{babel}

\usepackage[letterpaper,top=2cm,bottom=2cm,left=3cm,right=3cm,marginparwidth=1.75cm]{geometry}

\usepackage{amsmath}
\usepackage{amsfonts} 
\usepackage{graphicx}
\usepackage{authblk}
\usepackage{siunitx}
\usepackage{listings}
\usepackage{xcolor}

\definecolor{codegreen}{rgb}{0,0.6,0}
\definecolor{codegray}{rgb}{0.5,0.5,0.5}
\definecolor{codepurple}{rgb}{0.58,0,0.82}
\definecolor{backcolour}{rgb}{0.95,0.95,0.92}

\lstdefinestyle{mystyle}{
    backgroundcolor=\color{backcolour},   
    commentstyle=\color{codegreen},
    keywordstyle=\color{magenta},
    numberstyle=\tiny\color{codegray},
    stringstyle=\color{codepurple},
    basicstyle=\ttfamily\footnotesize,
    breakatwhitespace=false,         
    breaklines=true,                 
    captionpos=b,                    
    keepspaces=true,                 
    numbers=left,                    
    numbersep=5pt,                  
    showspaces=false,                
    showstringspaces=false,
    showtabs=false,                  
    tabsize=2
}

\usepackage[backend=biber, style=ieee]{biblatex}
\usepackage[colorlinks=true, allcolors=blue]{hyperref}

\DeclareMathOperator*{\argmin}{arg\,min}

\title{Hypersphere Optimization: Approximated Gradient-Based Binary Optimization on Smooth Manifold for Photonic Inverse Design}
\author[1,*]{Zhaocheng Liu}
\affil[1]{Meta Reality Labs, 9845 Willows Rd. Redmond, WA 98052}
\affil[*]{Corresponding author: Zhaocheng Liu, zhaocheng@fb.com}

\date{}  
  
\begin{document}

\maketitle

\begin{abstract}
Photonic inverse design typically seeks designs parameterized by binary arrays, where the values of each element correspond to the presence or absence of material at a particular point in space. Gradient-based approaches to photonic inverse design often include thresholding of non-binary optimization arrays; when the thresholding is sufficiently sharp, binary designs are obtained. However, difficulty can arise due to vanishing gradients for sharp thresholds, which cause optimization to stall. Here, we present hypersphere optimization, a new method of carrying out binarized photonic optimization on a high dimensional manifold. We numerically show that, on the manifold, upstream gradients from the objective function can smoothly transition a design from one nearly-binary array to another with minimal pixel flips. Our method is an approximate gradient-based optimization method for binary arrays that can be applied to various photonic inverse design problems and beyond. 
\end{abstract}

\section{Introduction}

Inverse design has attracted substantial attention in recent years, due to its ability to find unconventional photonic designs with extremely high performance. It has been successfully employed to a wide range of design problems, such as photonic integrated circuits \cite{piggott2015inverse}, diffractive optical elements \cite{colburn2021inverse, sell2017large, jiang2019global}, metasurfaces and metamaterials \cite{li2022empowering, liu2018generative, ma2018deep, ma2021deep}, nanoparticles and plasmonics\cite{malkiel2018plasmonic, wiecha2021deep, peurifoy2018nanophotonic, kudyshev2021machine}, and many more. Inverse designed components have been integrated into a variety of systems, including  sensing systems \cite{lin2022end, tseng2021neural}, optical communications\cite{yang2021inverse}, optical analog computing \cite{backer2019computational, hughes2018training}, and display optics \cite{li2022inverse}. At the same time, due to its ability to generate unintuitive structures, inverse design has been used to push the boundary of physics in areas such as nonlinear optics  \cite{hughes2018adjoint, raju2022maximized} and active components \cite{chung2020tunable}. 

There are several methods to parametrize a photonic structure in an optimization context. Two common approaches are shape parametrization and density-based parametrization. In shape optimization, photonic structures are represented by polygons and curves, while in density-based optimization, structures are represented by arrangements of binary pixels or voxels, with e.g. a pixel value corresponding to material composition at a particular point in space. In this paper, we will primarily focus on the latter, which involves optimization algorithms for binary arrays. 

Various methods have been proposed to address binary optimization in inverse design. Gradient-free optimization methods, such as genetic algorithms, initialize a population of designs and evolve the designs by emulating natural selection \cite{jin2019complex, wiecha2017evolutionary, di2022large}. It is often straightforward to ensure that designs are binary, and in principle, these methods can produce globally optimal designs. However, the downside is the vast computing resources required to evaluate the population of designs. This typically restricts such methods to problems with relatively fewer degrees of freedom (DOFs). On the other hand, gradient-based methods can address problems with vastly more degrees of freedom. For high-dimensional photonics design problems, gradient-based optimization is enabled by the adjoint variable method, which allows computation of gradients with respect to the degrees of freedom at the cost of a single simulation \cite{lalau2013adjoint, molesky2018inverse}. 


While gradient-based methods can efficiently traverse a large design space, there is a challenge in arriving at designs that are both binary and high-performance. In some sense, this is due to a fundamental incompatibility of gradient-based methods and the space of binary solutions. Gradient-directed updates to a design are real-valued and typically small, while binary designs differ by an integer number of pixels.

Density-based topology optimization comprises one set of methods that aim to address this challenge. To obtain binary designs, density-based topology optimization generally employs thresholding that sharpens as the optimization progresses, eventually forcing the design to assume binary values. These can be complemented with nonlinear constraints to produce designs exhibiting some minimum length scale \cite{jensen2011topology, eschenauer2001topology, hammond2021photonic}. However, typical binarizations schemes (e.g. sigmoid functions) are very sensitive to hyperparameters: changing the binarization hyperparameters can easily disrupt an optimization trajectory, and cause major regressions in the objective value. An alternative strategy to overcome the incompatibility is to generate binary designs from a latent vector and approximate the binary gradients using a straight through estimator (STE) \cite{schubert2022inverse, bengio2013estimating}. Since the estimated gradient is not strictly the gradient of the binarization process, photonic optimization using STE exhibits significant variation in the objective value \cite{yin2019understanding}. 

In this paper, we propose a new inverse design scheme called hypersphere optimization. Whereas typical inverse design methods seek binary designs – which can be thought of as the corners of a hypercube – we choose the hypersphere which circumscribes the hypercube as our optimization domain. The hypersphere is a smooth manifold, and as we will show, this imparts favorable characteristics on an optimization trajectory. We provide optimization examples in photonics and diffractive optics to demonstrate the applicability of our algorithm in inverse design. Compared to traditional topology optimization, hypersphere optimization always generates predictable optimization trajectory and does not require careful selection of binarization hyperparameters. It should be noted that this paper aims at unveiling the properties of hypersphere optimization and is not intended to show designs for specific purposes.

\begin{figure}
\centering
\includegraphics[width=0.7\textwidth]{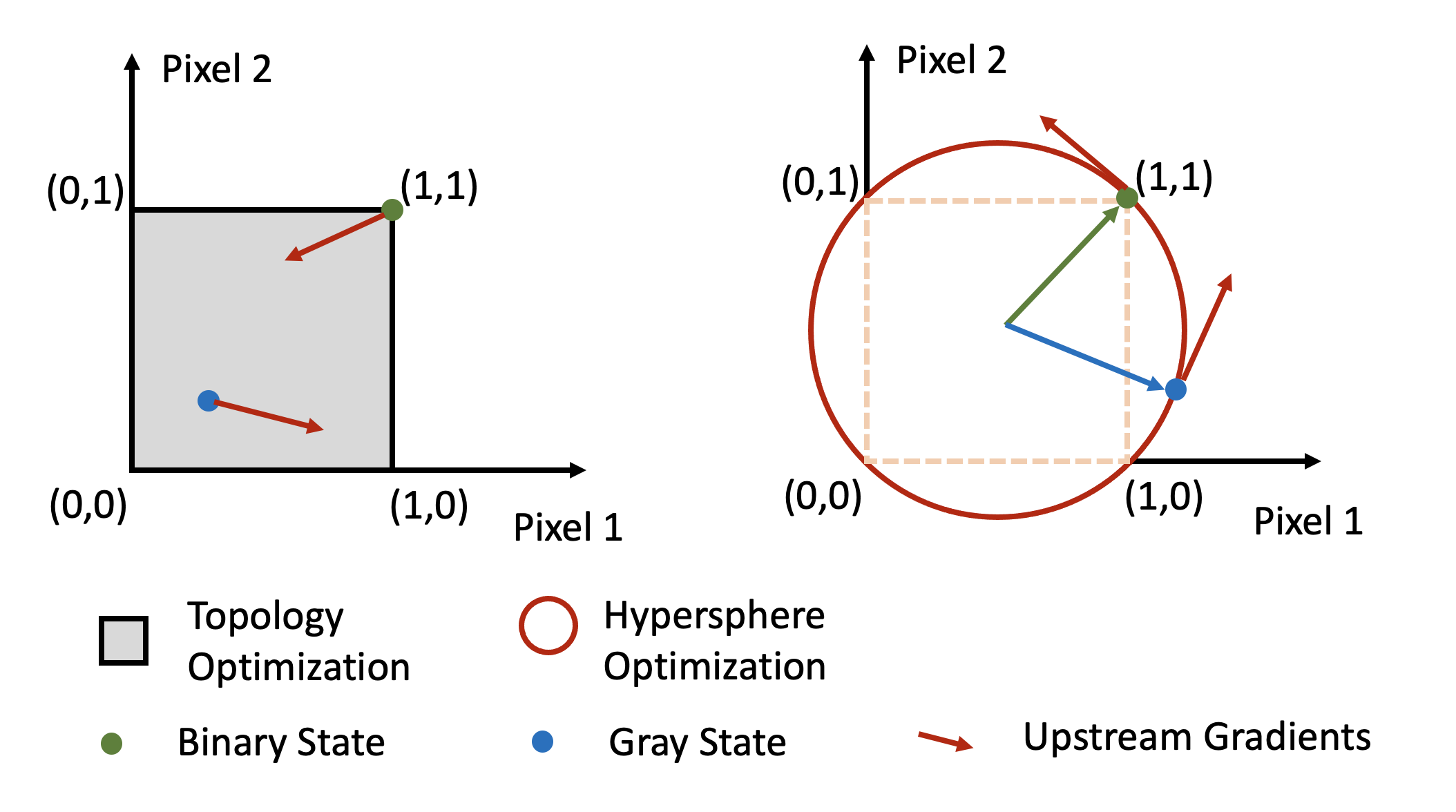}
\caption{ \textbf{Illustration of traditional topology optimization and hypersphere optimization.} In traditional topology optimization, the optimization domain is the hypercube bounded by 0 and 1. The desired binary states are on the corner of the cube. While in hypersphere optimization, the allowable optimization domain is the hypersphere circumscribing the hypercube. Binary states are inside the optimization domain. Given upstream gradients from the loss function, states in the hypercube optimization domain can point toward all directions, while upstream gradients on the hypersphere are tangential to the sphere. States on hypersphere can be transitioned smoothly along the geodesics between the two states guided by the upstream gradients. 
\label{Figure 1}
}
\end{figure}

\section{Optimization on Hypersphere}

\subsection{Hypersphere Optimization}
Figure 1 illustrates the conceptual difference between traditional topology optimization and hypersphere optimization. In both cases, we are interested in obtaining designs that are binary, which can be thought of as the corners of a hypercube. In topology optimization, the interior of the hypercube is traversed, typically starting at the center and then progressing to a corner. In hypersphere optimization, we consider a sphere which circumscribes the hypercube, and the optimization traverses the surface of the hypersphere.

The motivation for the hypersphere arises from the challenge of constructing a gradient-based optimization method for binary arrays. Specifically, the problem is as follows: given a binary array and upstream gradients from the loss function, determine the minimal pixels to flip in order to reduce the loss. Flipping minimal pixels is equivalent to finding a new binary state whose distance from the original is minimal, and whose direction is gradient-aligned. In the hypercube, the distance between two binary points is defined as the shortest connected edges between the two points. However, such distance is not a well-defined differentiable metric due to the sharp corners of the hypercube. On the hypersphere, however, the distance between any two binary states can be defined as the geodesic between the two points. The gradients are always tangential to the manifold surface. We can reduce the loss by moving a binary state to its neighboring binary state along the gradient direction, which is also along the shortest geodesic between the two states. Thus, optimizing on hypersphere can be thought of as flipping minimal pixels to minimize the loss function. 

We turn now to the issue of binarization. In hypersphere optimization, the desired binary states are inside of the optimization domain, in contrast to hypercube optimization where the binary states are on the corner of the optimization domain. State quantization inside the optimization domain can be more achievable than the quantization on the boundary. To illustrate this, we can think of binarization as pushing an unbinarized state (or gray state in Fig. 1) to the closest binary state. In hypercube optimization, a binarization scheme can squeeze a gray state to one of the corners of the hypercube. Once a binary state is locked, it is unlikely to move the stucked state out of the corner. In hypersphere optimization, since all binary states are defined on a smooth manifold, the state transition from a binarized state to an adjacent one suffers much less friction compared to the state transition in the hypercube. Given the gradients from the loss function, a binarized state can always keep transferring to its adjacent or close binarized states until the upstream gradient dies away and fails to beat the binarization friction. At such a point, we achieve a local optimum on the hypeshere. We will numerically demonstrate in later sections that optimization can still move forward even with strong binarization schemes.

\subsection{Algorithm Formulation }
We formulate a binary inverse design problem as follows. Given a $N$-dimensional binary vector $x_b \in \{0,1\}^N$  with elements corresponding to the pixels or voxels comprising a photonic device design, and a loss function $L(x)$ to be minimized, we search for an $x$ such that 

\begin{equation}
    x = \argmin(L(x_b))
\end{equation}

\noindent In contrast to objective functions in integer optimization problems, in inverse design the function $L(x)$ is a differentiable function on the continuous domain $x \in \mathbb{R}^N$ , and the derivative $\partial L/\partial x$  is well defined. 

\begin{figure}
\centering
\includegraphics[width=0.7\textwidth]{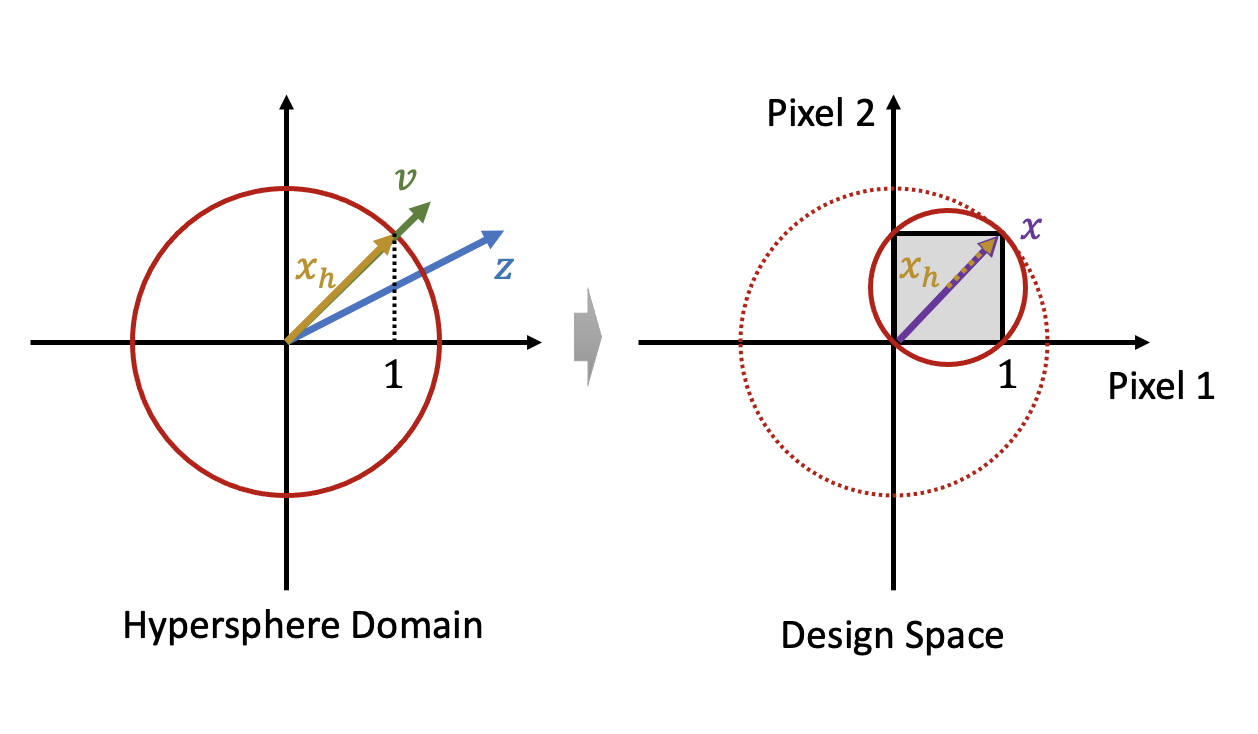}
\caption{\label{Figure 2} \textbf{Decode design vector from latent vector in hypersphere optimization.} (left) The hypersphere optimization domain is centered at origin with a radius of $\sqrt{N}$ . Given a latent vector $z$, it is first binarized to $v$ using hyperbolic tangent function. $v$ is then transformed to a vector $x_h$ that is strictly on the hypersphere. (right) The hypersphere state $x_h$ is mapped to design space by shifting and scaling
}
\end{figure}

Figure 2 summarizes the algorithm of hypersphere projection. The goal is to generate a binarized vector $x$ from a latent vector $z$ with the help of the hypersphere. Further, we require that the transform from $z$ to $x$ be differentiable, so that we can seamlessly backpropagate the upstream gradient from the loss function to the latent vector $z$ for differentiable optimization. Without loss of generality, we define the hypersphere centered at the origin: 

\begin{equation}
    \sum_i^N x_{hi} ^ 2 = R^2
\end{equation}

\noindent where $x_h$  denotes the points on the hypersphere and $R$ is the radius of the sphere. Since the $R$ does not affect the optimization landscape, for simplicity, we choose $R=\sqrt{N}$.  Given a latent vector $z$, we first binarize it towards binary state on hypersphere using hyperbolic tangent function: 

\begin{equation}
    v = \tanh{\beta z}
\end{equation}

\noindent where $\beta$ is a hyperparameter controlling the strength of binarization. The binarized state $v$ may not exactly lie on the sphere. We further normalize the vector on the sphere through the transformation:

\begin{equation}
    x_h = R \frac{v}{\|v\|}
\end{equation}

\noindent Next, we map $x_h$ to a vector $x$ in the design space for simulation. This can be done by shifting and scaling: 

\begin{equation}
    x = \frac{1}{2} (x_h + 1)
\end{equation}

\noindent The generated design vector $x$ will be evaluated by a differentiable simulation method. In optimization, we find the gradient of $L$ with respect to the latent vector $z$ and iteratively update $z$ by gradient descent. 

\subsection{Toy Example}

\begin{figure}
\centering
\includegraphics[width=0.5\textwidth]{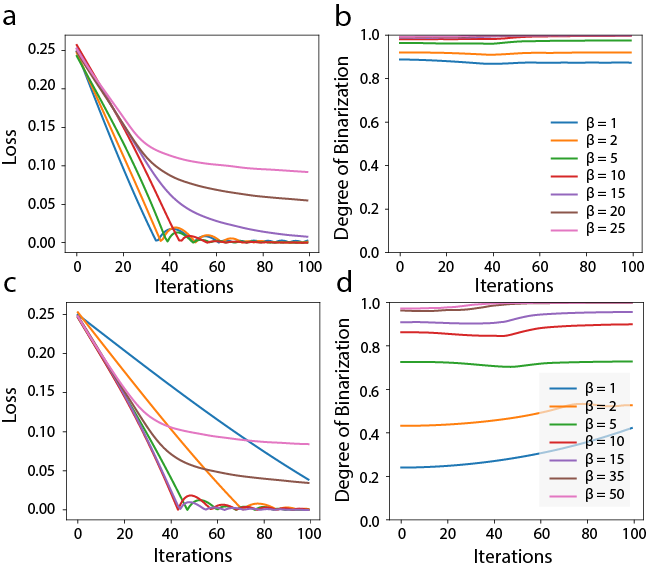}
\caption{\label{Figure 3} \textbf{Toy example of binarized hypersphere optimization and hypercube optimization.} (a–b) Loss and DOB variation in hypersphere optimization for different binary strength $\beta$.  controls the convergence rate of the optimization and the DOB of the optimized vector. The DOBs are sustained close to 1, indicating the vectors are mostly binarized througout the optimization. (c–d) Loss and DOB variation in hypercube optimization. The convergence does not present a consistent trend with respect to $\beta$. The variation of DOBs is also greatly affected by the choice of $\beta$.
}
\end{figure}

To test that hypersphere optimization can maintain a high binarization level during optimization, we compare it with hypercube optimization using a toy example. Assume we need to optimize a vector $x \in \mathbb{R}^{1024}$ where the mean of all the elements in the vector is required to be a constant $\alpha = 0.25$. We can formulate the loss function as:

\begin{equation}
    L(x) = \|\frac{1}{N}\sum_i x_i  - \alpha \|
\end{equation}

\noindent As discussed, we seek solutions where the elements of  $x$ are binary. As a success metric, we define degree of binarization (DOB) as:
\begin{equation}
    DOB = 2 \|\frac{1}{N} \sum_i x_i - \frac{1}{2} \|
\end{equation}

\noindent The DOB goes to 1 for binary images and converges to 0 for gray image with $x_i =0.5$ for all $i \in N$.  In both hypercube and hypersphere optimization, we initialize the optimizable latent vector $z$ by uniformly sampling each element $z_i$ from $-1$ to $1$. For hypersphere optimization, we transform/binarize $z$ using the scheme shown above. In hypercube optimization, we adopt the classical binarization scheme in traditional topology optimization, i.e., binarizing the latent vector to $x$ using sigmoid function:

\begin{equation}
    x = \frac{1}{1 + \exp(-\beta z)}
\end{equation}

\noindent Here the hyperparameter $\beta$ also represents the binarization strength.  In both toy examples, we optimize the latent vector using gradient descent in PyTorch with the same hyperparameters in the optimizer.

Figure 3 shows the loss and DOB of $x$ during the optimization of the two toy examples with different binarization strengths $\beta$. It should be noted that the value $\beta$  have  different impacts on the loss in the two examples. We only care about the trend of optimization with respect to $\beta$ regardless of the exact value of $\beta$. In hypersphere optimization (Fig. 3 (a–b)), the value of $\beta$  presents a clear impact on the optimization trajectory. A greater $\beta$ keeps the vector with higher DOB, while slows down the convergence of optimization. This coincides with our assumption that $\beta$ determines the resistances between state transitions and thus affect the speed of convergence. However, even with a small $\beta=5$, the algorithm sustains the DOB in a very high level (DOB $>0.95$), indicating the optimization happens on a mostly binarized space. In hypercube optimization (Fig. 3(c–d)), optimization trajectory is unpredictable given the choice of $\beta$. Both small or large $\beta$  can slow down the convergence rate. In addition, the DOB is largely dependent on $\beta$. This is one of the reasons that hyperparameter selection is crucial in traditional topology optimization for a high-performance binarized design. From this comparison, we can conclude that hypersphere optimization is able to backpropagate gradients to an almost binarized vector for optimization and shows predictable impact on the convergence of the optimization.

\section{Examples in Photonic Inverse Design}
To demonstrate the applicability of hypersphere optimization, we applied hypersphere optimization on various photonic inverse design problems. Like topology optimization, hypersphere optimization itself always generates structures with unreasonably minimum features. Luckily, we can directly use many topology optimization smoothing schemes. Here we choose to apply Gaussian blurring on the latent design $z$ with a pixel radius of $\sigma$, where $\sigma$  is defined as the variance of the Gaussian filter. The smoothed latent design is then binarized using hyperbolic tangent function, decoded to the corresponding design structure, and fed into a differentiable simulation function for the purpose of calculating the gradient of the latent design with respect to the loss function. Because the Gaussian blurring naturally incurs gray-scale states, we may need to increase $\beta$  when using filters with large $\sigma$. We should note that there may be more effective fabricability constraints methods on smooth manifolds that differ from the schemes in topology optimization. 

\begin{figure}
\centering
\includegraphics[width=1\textwidth]{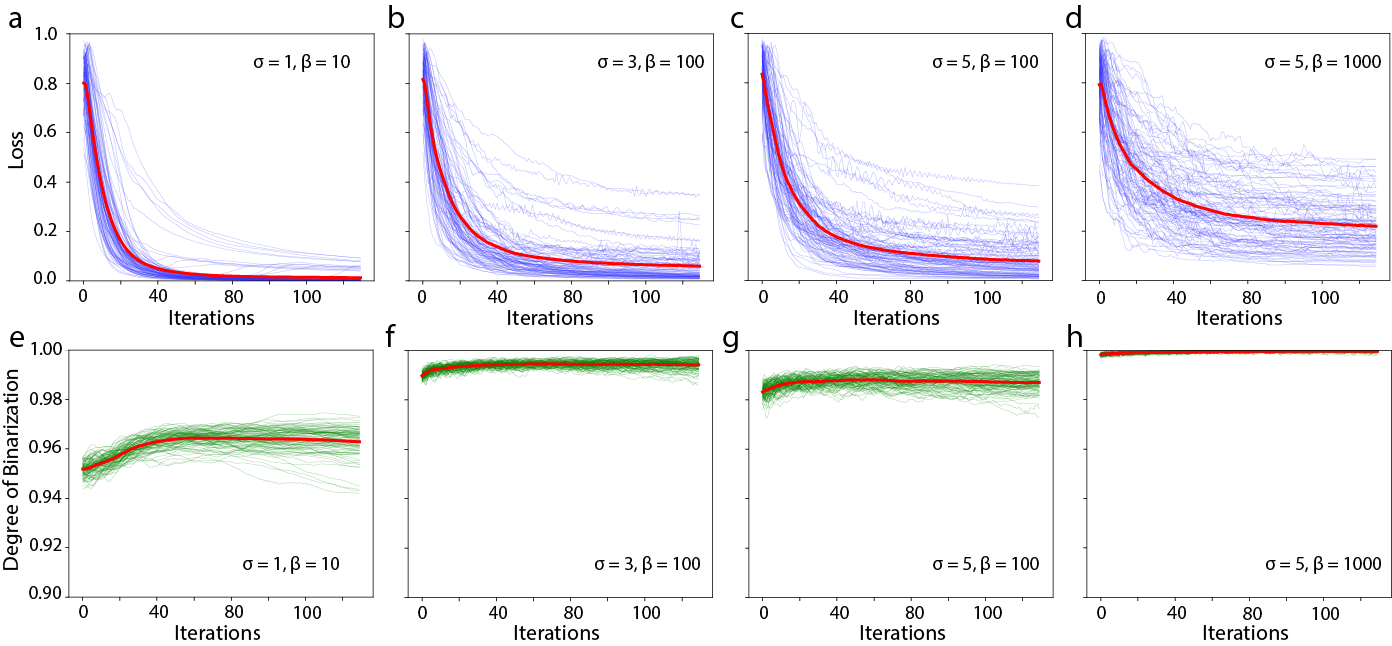}
\caption{\label{Figure 4} \textbf{Loss and DOB for waveguide bend optimization.} (a–d) Loss of the hypersphere optimization for waveguide bend with filter radius $\sigma=1, 3, 5, 5$, and binarization strength $\beta=10, 100, 100, 1000$, respectively. The convergence rate slows down with greater $\beta$  and $\sigma$, which corresponds to harder binarization strength and fewer design degrees of freedom. (e–h) DOB variations during the waveguide bend optimization. All DOBs throughout the optimization are close to 1. 
}
\end{figure}

\subsection{Waveguide Bend}

We first use a 90-degree waveguide bend design problem from Ceviche Challenge \cite{hughes2019forward, schubert2022inverse} as a hypersphere optimization example. The design region has a size of $\SI{1.6}{\micro\metre}$ by $\SI{1.6}{\micro\metre}$  and the width of input/output waveguide $\SI{0.4}{\micro\metre}$ . Our goal is to maximize the transmissions of the waveguide bend at $\SI{1270}{\nano\metre}$  and $\SI{1290}{\nano\metre}$. The loss function in the optimization is chosen as

\begin{equation}
    L = 1 - \frac{1}{2} (T_1 + T_2)
\end{equation}

To investigate the effect of hyperparameter $\beta$  and Gaussian filter radius $\sigma$ on the optimization trajectory, we picked four sets of combinations $(\beta, \sigma) = (10,1)$, $(100, 3)$, $(100, 5)$, and  $(1000, 5)$. For each combination, we carried out 100 optimizations with randomly initialized latent vectors. The initialization scheme is the same as the one in the  toy example.

The loss and DOB of the four optimization experiments are concluded in Figure 4(a) – 4(d). In all four cases, the loss value reduces in an almost monotonic way. With the increase of filtering radius $\sigma$, we observe worse performances on the converged designs. This can be understood by the fact that larger filter size results in less degrees of freedom in design, and thus the optimization scheme has a smaller chance to yield a good design. The interesting feature of hypersphere optimization is presented by the DOB throughout the optimization as shown in Fig. 4(f) – 4(h). Regardless of the choice of $\sigma$, the DOBs can maintain very high values given a properly chosen $\beta$. In the case where $\beta$ is set to 1000, it is surprising to see the consistent reduction of loss while the DOB keeps around 1. This means that gradient-based optimization can happen on nearly-binary design space using hypersphere optimization.

Figure 5(a) presents the best design out of 100 samples with $\beta =100$ and $ \sigma =5$. The corresponding electric fields at $\SI{1270}{\nano\metre}$ are shown on the right of the design. The transmissions at $\SI{1270}{\nano\metre}$ and $\SI{1290}{\nano\metre}$ are $T_1=98.88\%$ and $T_2=98.74\%$, respectively. Further, to investigate how well the design is binarized, we fully binarize the optimized design with a threshold of 0.5. We should note that this full binarization may note be necessary in practice because it changes the represented shape of the design on the meshgrid \cite{farjadpour2006improving}. The binarized structure and its simulated electric fields are shown in Fig. 5(b). Compared to the device before full binarization in Fig. 5(a), the changes of the topology and its electric fields are insignificant. The transmissions of the fully binarized devices are $97.81\%$ at the two wavelengths, marking a $1.07\%$ and $0.93\%$ performance drop compared to the structure before full binarization. In Figure 5(c), we summarize the transmissions of all optimized bends with different $\beta=100$ and $\sigma =5$ before and after full binarization.  The red line represents the ideal case where designs are binary and full binarization would not incur performance change. Most designs after full binarization do not exhibit performance drop. This indicates the optimized devices from hypersphere optimization have already been well binarized. Unless necessary, users do not need to manually adjust $\beta$ during the optimization to identify a binarized structure.

\begin{figure}
\centering
\includegraphics[width=0.6\textwidth]{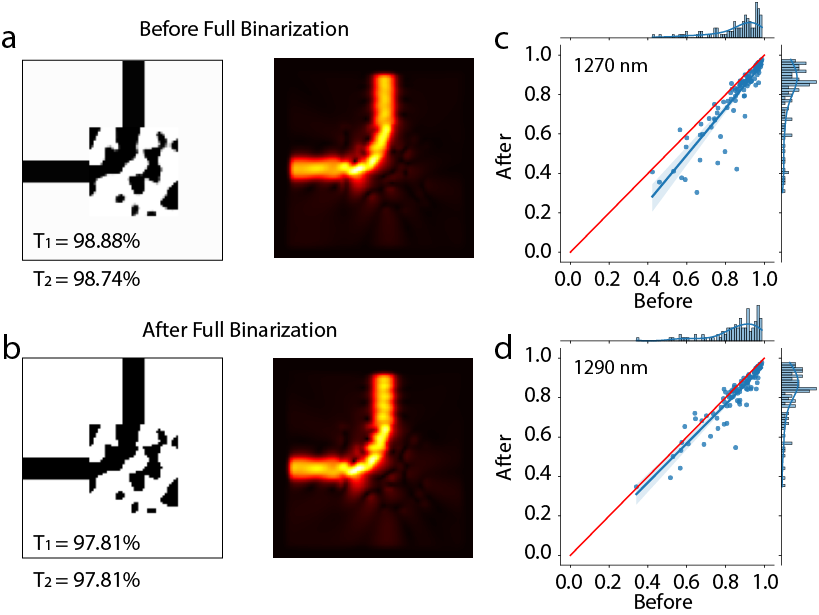}
\caption{\label{Figure 5} \textbf{Best waveguide bend design and full binarization test.} (a) Best design and its electric fields at $\SI{1270}{\nano\metre}$ among the 100 optimized structure with $\sigma=5$ and $\beta=100$. The transmittances at the two wavelengths are above $98\%$. (b) Fully binarized design from (a) and its electric fields at $\SI{1270}{\nano\metre}$. The transmittances reduce to $97.81\%$ at $\SI{1270}{\nano\metre}$  and $\SI{1290}{\nano\metre}$  wavelengths. (c–d) Transmission distributions for all 100 designs with $\sigma =5 $ and $\beta=100$ at (c) $\SI{1270}{\nano\metre}$ and (d) $\SI{1290}{\nano\metre}$ before and after full binarization. The red line indicates the ideal case where the designs are binary and full binarization does not incur performance change.
}
\end{figure}

\subsection{Mode Converter }

We further test our algorithm on inverse design for mode converters. Similar to the bend design example, we choose to optimize at two wavelengths $\SI{1270}{\nano\metre}$  and $\SI{1290}{\nano\metre}$ . The design region dimension and input/output waveguide size are the same as the bend example. The desired converter is required to convert fundamental mode to second order mode with maximum transmission. Figure 6(a) presents the best design of 100 randomly initialized optimizations with $\beta=100$ and $\sigma=5$. The electric field response is also presented on the right. The fully binarzed structure and their corresponding electric fields are shown in Figure 6(b). The most significant change of the device topology happens in the gray area in the center of the converter. The transmission of the fully binarized device are $T_1=95.90\%$ and $T_2=94.90\%$ at the two wavelengths, which have a reduction of $3.10\%$ and $4.25\% $compared to the structure transmissions before full binarization. We also summarize all the performance differences before and after the binarization for the 100 samples in Fig. 6(c). The transmissions of the 100 samples show similar distributions before and after full binarization. More examples with other $\beta$  and $\sigma$ combinations are included in Supplementary Information. 

\begin{figure}
\centering
\includegraphics[width=0.6\textwidth]{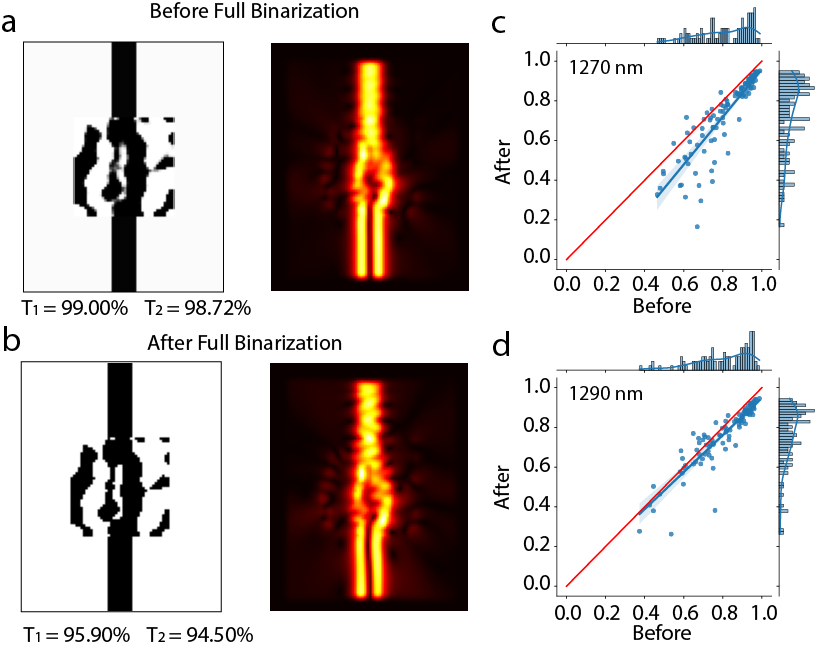}
\caption{\label{Figure 6} \textbf{Best mode converter design and binarization test.} (a) Best design and its electric fields at $\SI{1270}{\nano\metre}$ among the optimized structure with $\sigma=5$ and $\beta=100$. (b) Fully binarized design from (a) and its electric fields at $\SI{1270}{\nano\metre}$. (c–d) Transmission distributions for all 100 designs with $\sigma=5$ and $\beta=100$ at (c) $\SI{1270}{\nano\metre}$ and (d) $\SI{1290}{\nano\metre}$ before and after full binarization. 
}
\end{figure}

\section{Diffractive Optical Element}
To further test the generalizability of our algorithm, we applied hypersphere optimization to inverse design of diffractive optical elements (DOEs). The design configuration is shown in Figure 7(a). The thickness of the design region is $\SI{0.9}{\micro\metre}$ , and periodicity of the DOE is $\SI{5}{\micro\metre}$  by $\SI{5}{\micro\metre}$ . The refractive of the substrate is set to $n_{sub}=1.55$. $x$-polarized incident light  with a wavelength $\lambda = \SI{0.94}{\micro\metre}$ is excited from the substrate and is diffracted into air. There are a total 11 supported diffraction orders in $x$-direction and 89 orders in all directions. The maximum diffraction angle in 1D is about $\theta_{max}=70^{\circ}$. Our goal is to optimize the uniformity of all the diffracted lights. The material of the DOE is initially set to be $n_{mat}=1.55$ and the gaps between the DOE pattern is $n_{gap}=1$. We should note that the preset indices of DOE $n_{mat}$, $n_{gap}$ may not support such an extreme uniformity requirement, and thus we also treat them as optimizable parameters. 

\begin{figure}
\centering
\includegraphics[width=0.55\textwidth]{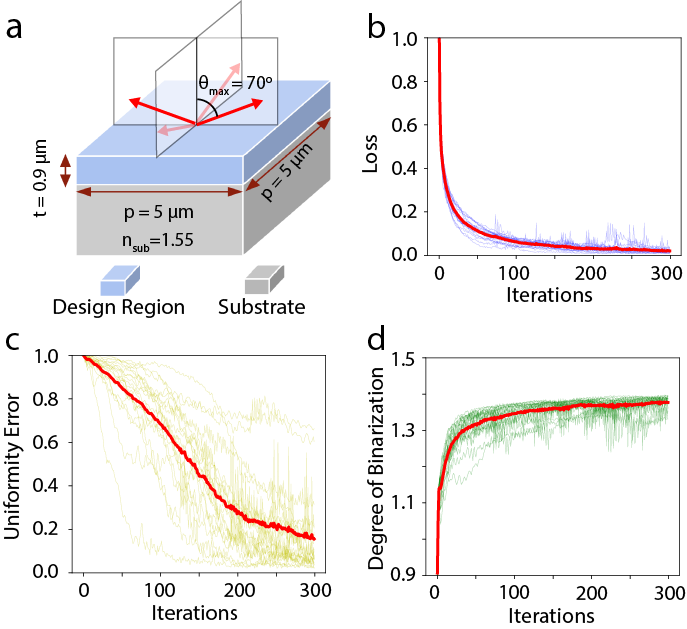}
\caption{\label{Figure 7} \textbf{DOE design configuration and optimization. (a) DOE optimization configuration.} The index of the substrate is $n_{sub}=1.55$ at the target incident wavelength $\SI{0.94}{\micro\metre}$. The periodicity and the thickness of the DOE are  $p= \SI{5}{\micro\metre}$ and $t = \SI{0.94}{\micro\metre}$, respectively.  We initialize the indices of the material of the DOE $n_{mat}=1.55$ and the gaps between the structure  $n_{gap}=1$. There are 11 diffraction orders in $x$-direction and 89 orders in all directions. The 1D maximum diffraction angle is $\theta_{max}=70^{\circ}$. The objective is to optimize all 89 diffracted lights to achieve a uniform intensity distribution. (b–d) Losses, uniformity errors, and DOBs of all 20 designs during the optimization. The DOBs exceeds 1, indicating the binarized structures have indices $n_{mat}>1.55$ and $n_{gap}<1$.
}
\end{figure}

In optimization, we want to minimize the uniformity error:
\begin{equation}
    U_{err} = \frac{I_{max} - I_{min}}{I_{max} + I_{min}}
\end{equation}

\noindent where $I_{max}$  and $I_{min}$ are the maximum and minimum intensities of all the 89 diffracted lights. However, treating $U_{err}$ as the loss function in optimization is not ideal because it only optimizes 2 (maximum and minimum) beams at each iteration. Instead, we choose to minimize the loss function:

\begin{equation}
    L_{DOE} = \frac{1}{M} \sum_i (I_i - I^\star) ^2 
\end{equation}

\noindent where $M=89$ is the number of diffraction orders and $I^\star$ is the average intensity of all diffracted lights. We enforce the design to be centrosymmetric as the desired diffractive patterns in the far field are spatially symmetric. We randomly initialize 20 latent vectors for optimization. The binarization strength $\beta$ is chosen to be 10. 

The loss, uniformity, and DOB variation of the 20 optimizations are shown in Fig. 7(b–d). The losses of all designs consistently reduce during the optimization, and the uniformity errors drop after the losses are sufficiently small. However, the DOB variation in Fig. 7(d) exceeds 1 all the way in the optimization. This indicates the design is almost binarized but has maximum binary value above 1 and minimum value below 0, which corresponds to $n_{mat}>1.55$ and $n_{gap}<1$. In hyperparameter optimization, the change of the DOE indices is majorly attributed  to the choice of $\beta$.  The $\beta$ in the optimization is large enough to force most of entries in $v$ on the hypersphere approaches binary values, but it still allows, a tiny number of $v_i$’s to move away from the binary values. This portion of un-biarized $v_i$’s lead the optimized state land on the hypersphere region hovering away from the hypercube (e.g. the gray state on hypersphere optimization in Fig. 1), causing the optimized structure converging to values above 1 and below 0. In the meantime, diffractive structure naturally prefers high index contrast. Thus, we observe that the index of the structure is also tuned in hypersphere optimization.

Figure 8 shows the best and worst designs and their corresponding diffraction intensity distributions. In the two best designs in Fig. 8(a) and 8(b), the indices of the DOEs are $n_{mat}=1.65$, $n_{gap}=0.86$, and $n_{mat}=1.67$, $n_{gap}=0.82$, respectively. The  diffracted lights are uniformly distributed across all the diffraction orders with both the uniformity errors below $4\%$. In the two worst designs in Fig. 8(c) and 8(d), we observe a similar index contrast of the DOE patterns. Although the uniformity errors are significant, the errors majorly come from the zeroth order diffraction. The feature of cooptimize indices of material in hypersphere optimization may provide new optimization strategies for diffractive optics inverse design. With that being said, it is always feasible to increase $\beta$  to force the maximum and minimum value of the topology pixel to be 0 and 1. 

\begin{figure}
\centering
\includegraphics[width=0.75\textwidth]{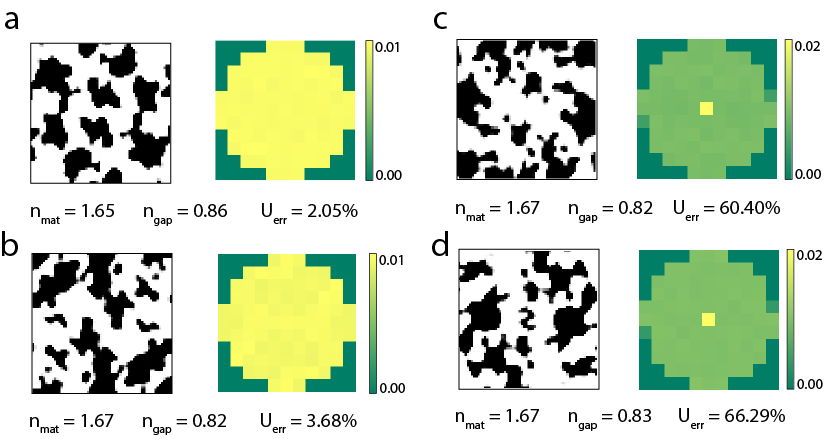}
\caption{\label{Figure 8} \textbf{Best and worst DOE design examples.} (a–b) Two best designs out of the 20 random initialized optimization. The black and white regions represent the material $n_{mat}$ and gaps $n_{gap}$. The diffraction intensity distributions of the two designs are shown on the right of each design. The uniformity errors of the two examples are $2.05\%$ and $3.68\%$. (c–d) Two worst designs and their far field intensity distributions from the 20 samples. The high uniformity errors are majorly contributed by the center diffractive beam. 
}
\end{figure}

\section{Conclusion and Outlook }
In summary, we have proposed a new optimization scheme that carries out binary optimization  on a hypersphere for photonic inverse design. Due to the well-defined differentiability on the smooth manifold, binary states can be seamlessly transitioned among each other on the hypersphere. This feature allows us to back propagate gradients on the hypersphere and, with a proper binarization scheme, enables the gradient-based optimization for almost binarized variables. The binarization strength in hypersphere optimization has a straightforward impact on the optimization trajectory and DOB. Compared to traditional topology optimization, our algorithm can yield a well-binarized design with little requirement on adjusting the binarization hyperparameters during optimization. We have numerically demonstrated that hypersphere optimization is applicable to the inverse design problems for photonic structures and diffractive components. 

Hypersphere optimization shows interesting optimization capabilities for binary photonic structure, but we believe more benchmarks and investigation are required to fully unveil the performance and mechanisms of the method. In our examples, we assumed the loss function defined on the hypersphere is sufficiently smooth and the projected gradients on the hypersphere are nonzero. The assumption may need rigorous investigation for different physical processes to determine the broader applicability of the algorithm. On the other hand, many features of hypersphere optimization have not been thoroughly utilized. For instance, since we have defined distances between two binary vectors on the hypersphere, we may locate the closest feasible designs \cite{schubert2022inverse} that satisfy fabrication rules with minimum performance drop, to effectively constrain geometric features of the design. Another application of such well-defined distance metric may be data-efficient sampling for machine learning design methods \cite{ma2019probabilistic, liu2018training, so2020deep}, because strongly related design data can be clustered on the same region on the hypersphere.  We hope hypersphere optimization could provide a new perspective on inverse design, and inspire new optimization strategies to mitigate challenges in binary optimization problems in general.

\section*{Acknowledgement}
We thank Martin F. Schubert for his thoughtful discussion on the technical details and his suggestions and help that greatly improved the manuscript. 

\newpage 
\printbibliography

\newpage 

\section*{Supporting Information}
\setcounter{figure}{0}

\makeatletter 
\renewcommand{\thefigure}{S\@arabic\c@figure}
\makeatother

\begin{figure}[!htb]
\centering
\includegraphics[width=1\textwidth]{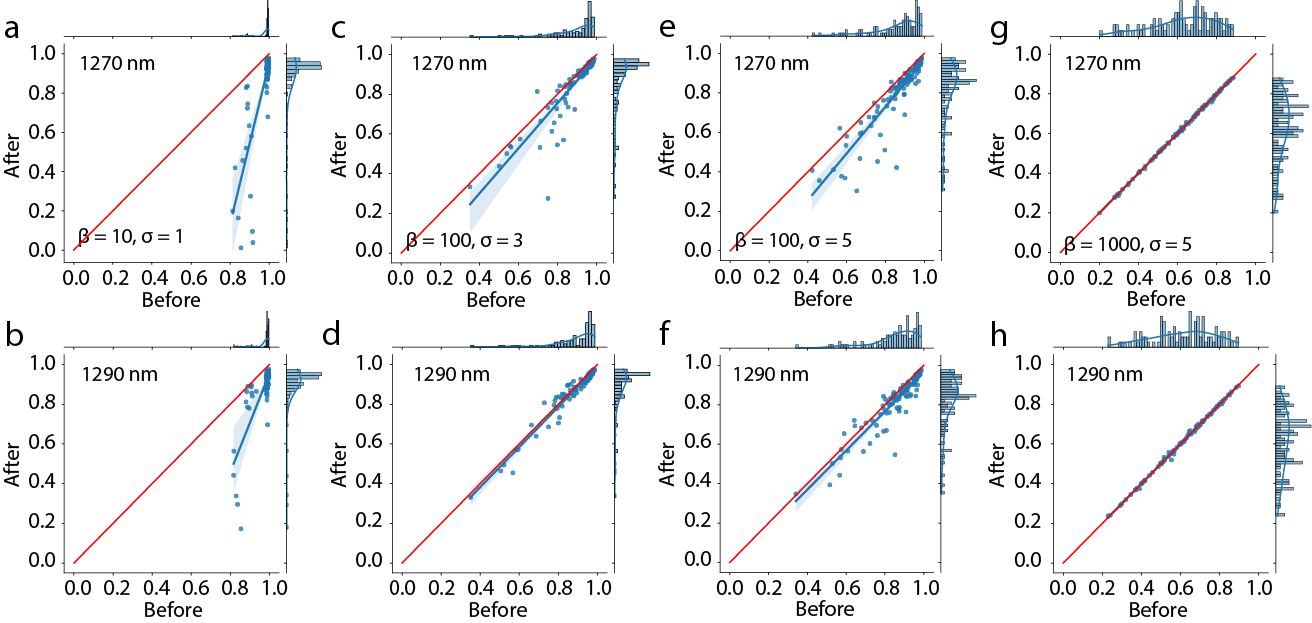}
\caption{\label{Figure 8} \textbf{Binarization test for all waveguide bend designs.}  (a–h) Transmission distributions for 100 designs with all  and  combinations at 1270 nm and 1290 nm before and after full binarization. The plot in Figure 4 $(\beta=100, \sigma=5)$ is also included for comparison. 
}
\end{figure}

\begin{figure}[!htb]
\centering
\includegraphics[width=1\textwidth]{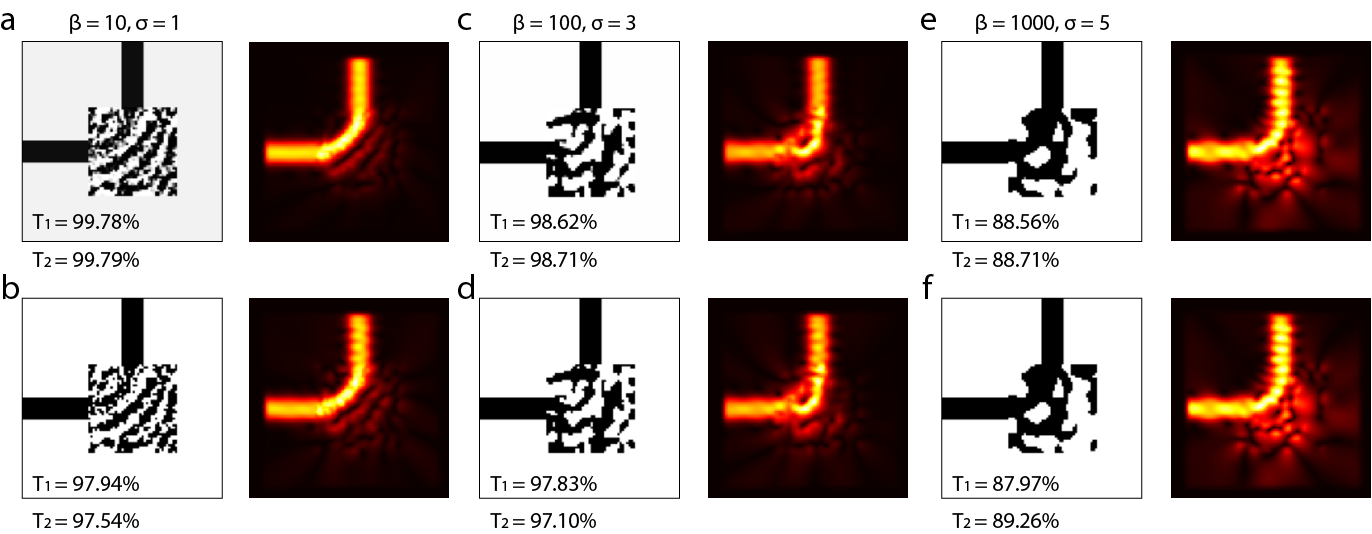}
\caption{\label{Figure 8} \textbf{Best waveguide bend designs.} (a–c) Best waveguide bend designs and the simulated electric fields at 1270 nm for optimization with $\sigma =$1,3,5 and $\beta=10$, 100, 1000 , respectively. (d–e) The fully binarized designs form (a–c) and the corresponding electric fields at 1270 nm. 
}
\end{figure}

\begin{figure}[!htb]
\centering
\includegraphics[width=1\textwidth]{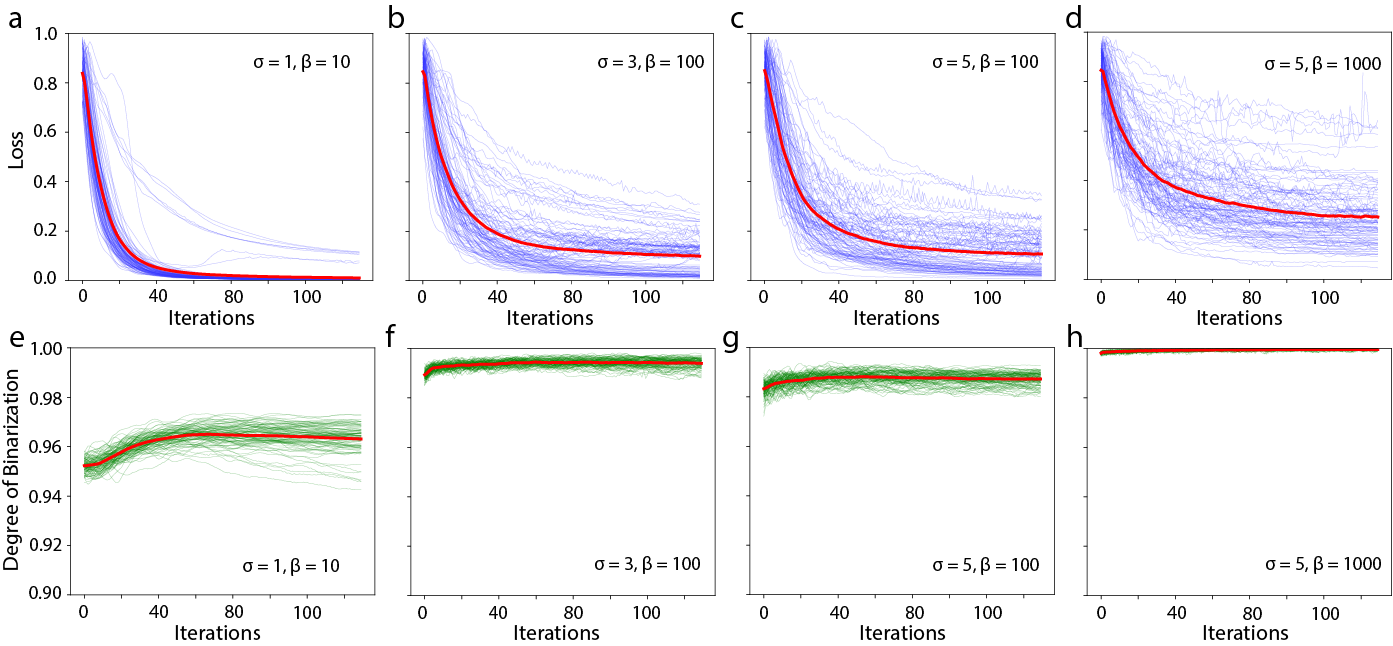}
\caption{\label{Figure 8} \textbf{Loss and DOB for mode converter optimization.} (a–d) Loss of the hypersphere optimization for converter with filter radius $\sigma =1$, 3, 5, 5, and binarization strength $\beta=10$, 100, 100, 1000, respectively. (e–h) DOB variations during the waveguide bend optimization. 
}
\end{figure}

\begin{figure}[!htb]
\centering
\includegraphics[width=1\textwidth]{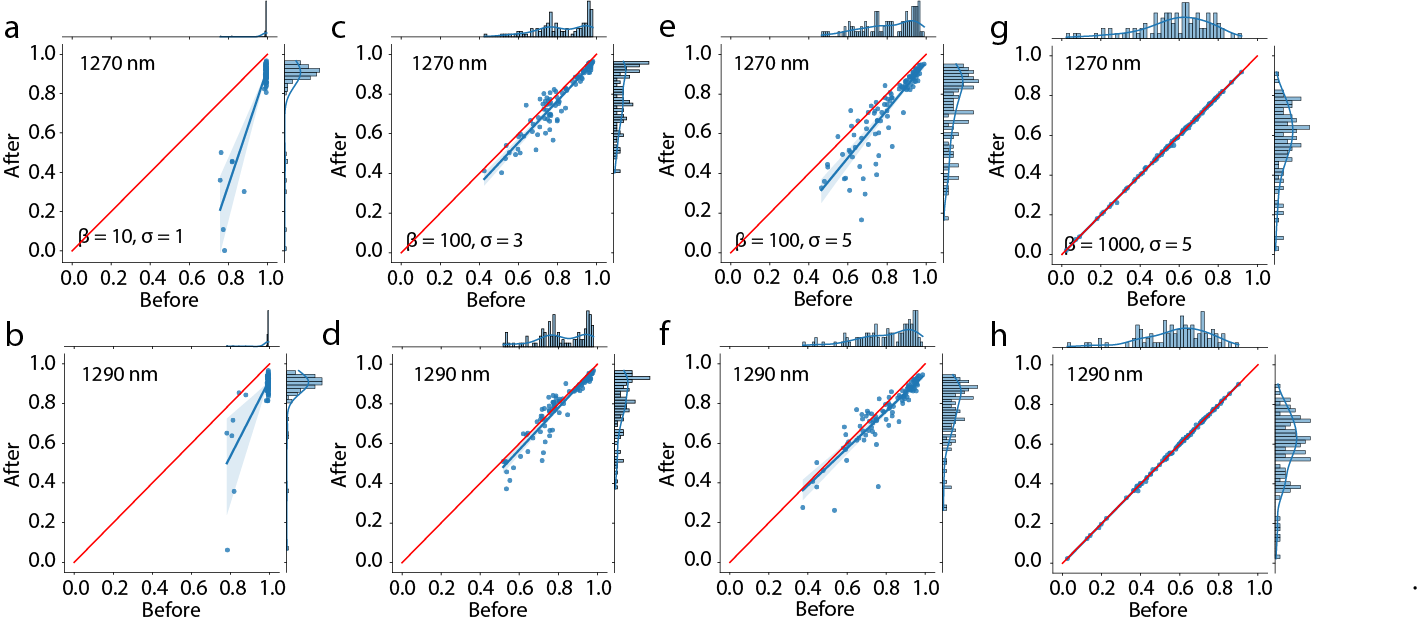}
\caption{\label{Figure 8} \textbf{Binarization test for all mode converter designs.}  (a–h) Transmission distributions for 100 designs with all  and  combinations at 1270 nm and 1290 nm before and after full binarization, respectively. The plot in Figure 5 $(\beta=100, \sigma=5)$ is also included for comparison. 
}
\end{figure}

\begin{figure}[!htb]
\centering
\includegraphics[width=1\textwidth]{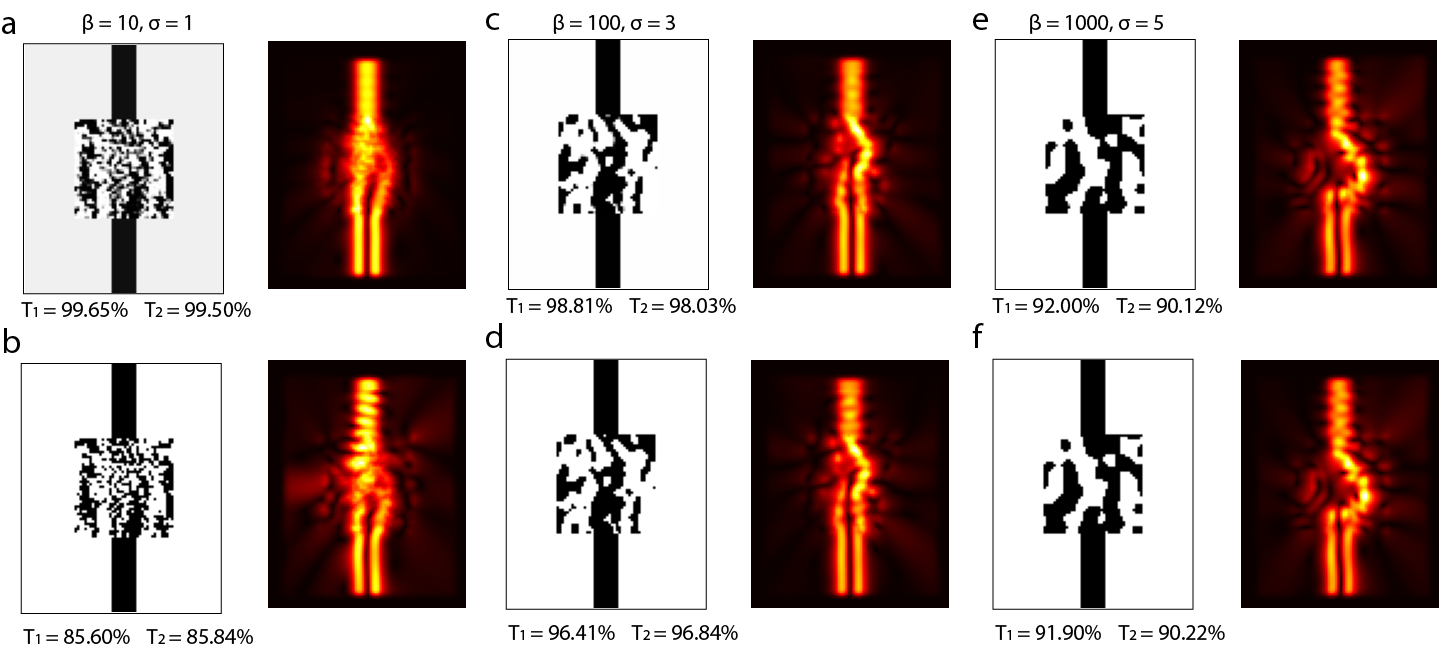}
\caption{\label{Figure 8} \textbf{Best mode converter designs.} (a–c) Best mode converter designs and the simulated electric fields at 1270 nm for optimization with $\sigma=1$,3,5 and $\beta=10$, 100, 1000 , respectively. (d–e) The fully binarized designs form (a–c) and the corresponding electric fields at 1270 nm. 
}
\end{figure}

\begin{figure}[!htb]
\centering
\includegraphics[width=1\textwidth]{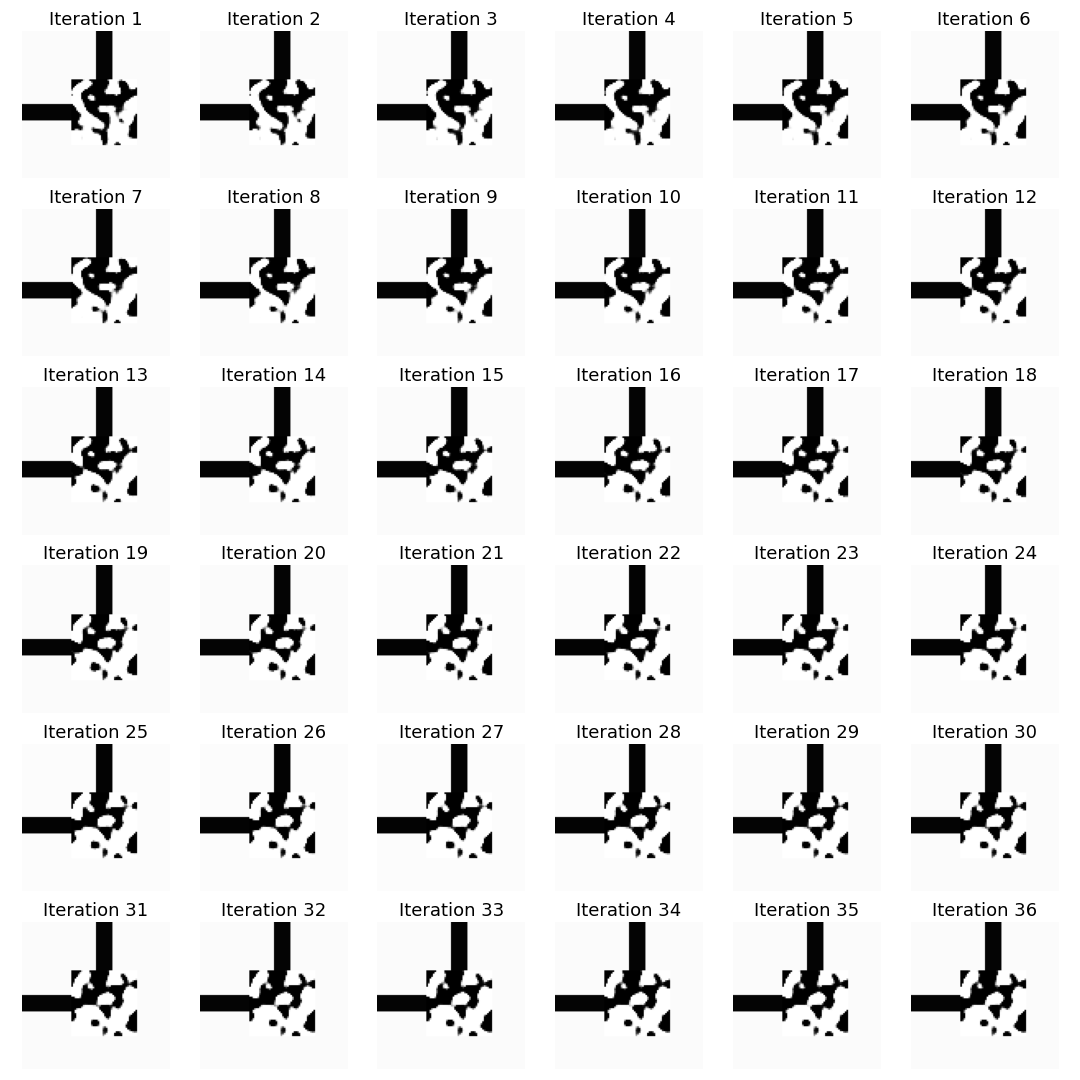}
\caption{\label{Figure 8} \textbf{Figure S6 | Topology variations during hypersphere optimization example 1.} First 36 topology changes during the optimization process in the example Figure 5 with $\beta=100$. Only tiny region of the topology of the waveguide bend varies in each iteration. }
\end{figure}

\begin{figure}[!htb]
\centering
\includegraphics[width=1\textwidth]{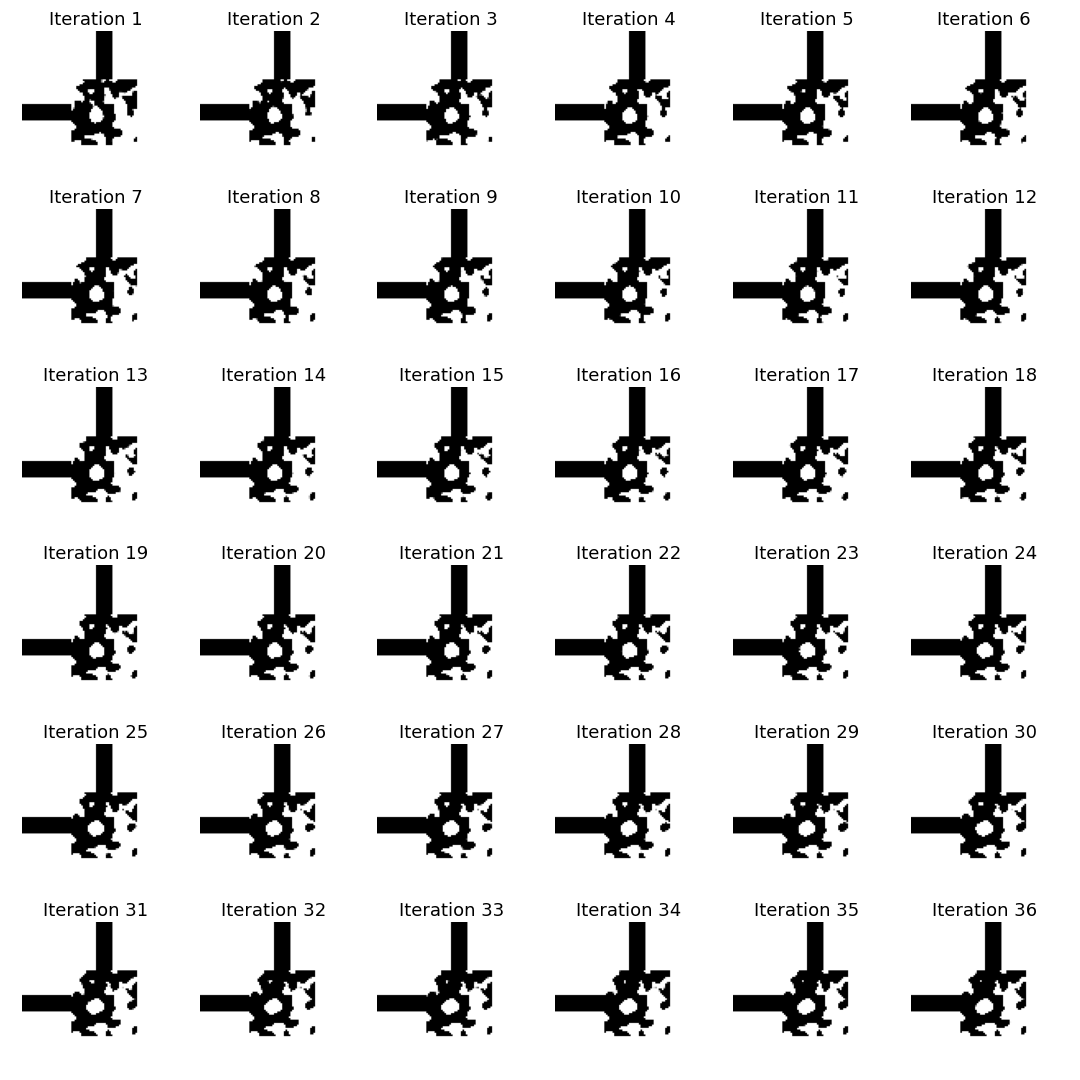}
\caption{\label{Figure 8} \textbf{Topology variations during hypersphere optimization example 2.}  First 36 topology changes during the optimization process in the example Figure 2(e) with $\beta=1000$. The variation of the topology of the waveguide bend is almost pixelwise. 
}
\end{figure}

\clearpage
\begin{lstlisting}[language=Python, caption={PyTorch implementation of hypersphere projection}]
import torch
import numpy as np
    
def hyperphsere_projection(z: torch.Tensor, beta: float):
    ''' Function that transforms latent design z to a design x
        with the binarization strength beta
        Args:
            z: the latent design. Shape (Nx, Ny)
            beta: binarization hyperparameter 
    '''
    
    # step 1: binarization to vector v (Eq. 3)
    v = torch.tanh(beta * z)  
    
    # step 2: map  v strictly on hypersphere xh (Eq. 4)
    Nx, Ny = z.shape
    N = Nx * Ny
    radius = np.sqrt(float(N))
    norm = torch.sqrt(torch.sum(torch.abs(v)**2)) 
    scale = norm / radius
    xh = v / scale 
    
    # step 3: transform to design space (Eq. 5)
    x = xh / 2 + 0.5
    return x

\end{lstlisting}

\end{document}